\documentclass[12pt]{article}
\usepackage{amsmath,amssymb,latexsym,epsf,epsfig,graphicx,tikz}

\begin{document}

\newcommand{\prt}{\partial}
\newcommand{\II}{\mbox{${\mathbb I}$}}
\newcommand{\CC}{\mbox{${\mathbb C}$}}
\newcommand{\RR}{\mbox{${\mathbb R}$}}
\newcommand{\QQ}{\mbox{${\mathbb Q}$}}
\newcommand{\ZZ}{\mbox{${\mathbb Z}$}}
\newcommand{\NN}{\mbox{${\mathbb N}$}}
\newcommand{\DD}{\mbox{${\mathbb D}$}}
\newcommand{\PP}{\mbox{${\mathbb P}$}}
\def\G{\mathbb G}
\def\UU{\mathbb U}
\def\S{\mathbb S}
\def\T{\mathbb T}
\def\KK{\mathbb K}
\def\J{\mathbb J}
\def\tT{\widetilde{\mathbb T}}
\def\HH{\mathbb H}
\def\tS{\widetilde{\mathbb S}}
\newcommand{\mB}{{\mathbb B}}
\newcommand{\mA}{{\mathbb A}}
\newcommand{\mC}{{\mathbb C}}
\newcommand{\mM}{{\mathbb M}}
\newcommand{\mJ}{{\mathbb J}}
\newcommand{\ma}{{\mathbb a}}

\def\V{\mathbb V}
\def\tV{\widetilde{\mathbb V}}
\newcommand{\D}{{\mathbb D}}
\def\hint{H_{\rm int}}
\def\R{{\cal R}}

\newcommand{\rd}{{\rm d}}
\newcommand{\diag}{{\rm diag}}
\newcommand{\cU}{{\cal U}}
\newcommand{\K}{{\mathcal K}}
\newcommand{\cP}{{\cal P}}
\newcommand{\dQ}{{\dot Q}}
\newcommand{\dS}{{\dot S}}
\newcommand{\dW}{{\dot W}}
\newcommand{\W}{{\mathcal W}}

\newcommand{\pnf}{P^N_{\rm f}}
\newcommand{\pnb}{P^N_{\rm b}}
\newcommand{\hnf}{P^Q_{\rm f}}
\newcommand{\hnb}{P^Q_{\rm b}}

\newcommand{\ph}{\varphi}
\newcommand{\phd}{\widetilde{\varphi}} 
\newcommand{\phs}{\varphi^{(s)}}
\newcommand{\phb}{\varphi^{(b)}}
\newcommand{\phds}{\widetilde{\varphi}^{(s)}}
\newcommand{\phdb}{\widetilde{\varphi}^{(b)}}
\newcommand{\vt}{\vartheta}
\newcommand{\lambdad}{\widetilde{\lambda}}
\newcommand{\tx}{\widetilde{x}} 
\newcommand{\td}{\widetilde{d}} 
\newcommand{\etat}{\widetilde{\eta}}
\newcommand{\phl}{\varphi_{i,L}}
\newcommand{\phr}{\varphi_{i,R}}
\newcommand{\phz}{\varphi_{i,Z}}
\newcommand{\mum}{\mu_{{}_-}}
\newcommand{\mup}{\mu_{{}_+}}
\newcommand{\mupm}{\mu_{{}_\pm}}
\newcommand{\muv}{\mu_{{}_V}}
\newcommand{\mua}{\mu_{{}_A}}
\newcommand{\wt}{\hat{t}}

\def\a{\alpha}
 
\def\A{\mathcal A} 
\def\H{\mathcal H} 
\def\U{\mathcal U} 
\def\E{\mathcal E} 
\def\C{\mathcal C} 
\def\L{\mathcal L} 
\def\M{\mathcal M} 
\def\O{\mathcal O}
\def\I{\mathcal I}
\def\Z{\mathcal Z} 
\def\der{\partial }
\def\mis{{\frac{\rd k}{2\pi} }}
\def\ri{{\rm i}}
\def\xt{{\widetilde x}}
\def\ft{{\widetilde f}}
\def\gt{{\widetilde g}}
\def\qt{{\widetilde q}}
\def\tt{{\widetilde t}}
\def\tmu{{\widetilde \mu}}
\def\prt{{\partial}}
\def\tr{{\rm Tr}}
\def\inc{{\rm in}}
\def\out{{\rm out}}
\def\Li{{\rm Li}}
\def\e{{\rm e}}
\def\eps{\varepsilon}
\def\k{\kappa}
\def\v{{\bf v}}
\def\ebf{{\bf e}}
\def\abf{{\bf A}}
\def\lb{\Omega_{{}_{\rm LB}}}
\def\rlb{\rangle_{{}_{\rm LB}}}
\def\hlb{\H_{{}_{\rm LB}}}

\pagestyle{empty}
\rightline{March 2020}

\bigskip 

\begin{center}
{\Large\bf Entropy Production in Systems with\\ Spontaneously Broken Time-Reversal}
\\[2.1em]

\bigskip

{\large Mihail Mintchev}\\ 
\medskip 
{\it  
Istituto Nazionale di Fisica Nucleare and Dipartimento di Fisica, \\ 
Universit\`a di Pisa, Largo Pontecorvo 3, 56127 Pisa, Italy}
\bigskip 

{\large Paul Sorba}\\ 
\medskip 
{\it  
Laboratoire d'Annecy-le-Vieux de Physique Th\'eorique, 
CNRS, \\ Universit\'e de Savoie,   
BP 110, 74941 Annecy-le-Vieux Cedex, France}
\bigskip 
\bigskip 
\bigskip 

\end{center}



\begin{abstract}

We study the entropy production in non-equilibrium quantum systems without dissipation, which is generated 
exclusively by the spontaneous breaking of time-reversal invariance. Systems which preserve the total energy and 
particle number and are in contact with two heat reservoirs are analysed. Focussing on point-like interactions, 
we derive the probability distribution induced by the entropy production operator. 
We show that all its moments are positive in the zero frequency limit. The analysis covers both Fermi and Bose statistics.

\end{abstract}

\vskip 5 truecm 

\begin{footnotesize}

\centerline {\it To appear in the Proceedings of the MPHYS10 meeting, Belgrade, Serbia, September 2019.}

\end{footnotesize}

\vfill
\newpage
\pagestyle{plain}
\setcounter{page}{1}

\section{Introduction}

\begin{figure}[h]
\begin{center}
\begin{picture}(600,22)(43,325) 
\includegraphics[scale=0.87]{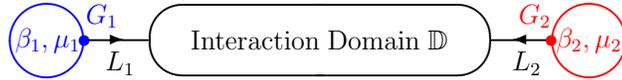}
\end{picture} 
\end{center}
\caption{System connected to two heath reservoirs.} 
\label{fig1}
\end{figure}

We investigate below non-equilibrium quantum systems with infinite degrees of freedom, where the time reversal 
symmetry $T$ is {\it spontaneously} broken. In particular, we explore the impact of the spontaneous 
$T$-breaking on the entropy production, 
which represents the key indicator for the departure from equilibrium. The presentation summarises our previous 
results \cite{MSS-17, MSS-18} and pursues further our study of the entropy production in systems with the structure shown in Fig. \ref{fig1}. 
One has two heat reservoirs $R_i$ with (inverse) 
temperatures $\beta_i\geq 0$ and chemical potentials $\mu_i$, which communicate via the gates $G_i$ 
with the interaction domain $\D$. 
Particles are emitted and absorbed by the reservoirs $R_i$ trough the gates $G_i$ and 
propagate along the leads $L_i$ to the  
interaction domain $\D$. The interaction in $\D$ drives 
the system away from equilibrium. The capacity of $R_i$ is assumed 
large enough so that the particle emission and absorption 
processes do not change their parameters. 

Systems of this type in one space dimension are widely investigated. They are successfully applied 
for studying the transport properties of quantum wire junctions \cite{kf-92}-\cite{bcm-09}, quantum Hall edges \cite{AS-08} and 
recently are also engineered in laboratory by ultracold Bose gases \cite{BDZ-08}-\cite{CS-16}. 
The remarkable precision reached in such experiments 
allows to explore fundamental aspects of non-equilibrium many-body quantum physics. 

In the next section we establish some general universal features of non-equilibrium systems, which 
communicate with two heat reservoirs. Afterwards, in section 3 we provide a microscopic picture 
for the entropy production in such systems and propose a quantum field theory framework for the derivation of 
the probability distribution generated by the entropy production operator. In section 4 we illustrate 
this approach at work exploring two examples - the fermionic and bosonic Schr\"odinger junctions with point-like 
interaction. The last section is devoted to the conclusions. 

\bigskip 

\section{Spontaneous breaking of time-reversal invariance} 

Quantum systems in contact with two heat reservoirs exhibit in general a complex behaviour. In order to 
simplify the picture, we assume in what follows the conservation of the particle number and the total energy 
\begin{equation}
N=\int_{G_1}^{G_2} \rd x\, j_t(t,x) \, , \quad H=\int_{G_1}^{G_2} \rd x\, \vt_t(t,x) \, , 
\label{g1}
\end{equation} 
$j_t$ and $\vt_t$ being the particle and energy densities. In other words there is no particle and 
energy dissipation. Let $j_x$ and $\vt_x$ be the associated local conserved currents which obey the continuity equations 
\begin{eqnarray}
\der_t j_t(t,x) - \der_x j_x(t,x)= 0\, , 
\label{g2} \\
\der_t \vt_t(t,x) - \der_x \vt_x(t,x)= 0\, .  
\label{g3} 
\end{eqnarray} 
Combining (\ref{g1}) with (\ref{g2},\ref{g3}) and taking into account the orientation of the leads $L_i$ one finds 
\begin{eqnarray}
{\dot N}= 0\quad  \Rightarrow \quad j_x(t,G_1) + j_x(t,G_2)=0\, , \; 
\label{g4} \\
{\dot H}= 0\quad  \Rightarrow \quad \vt_x(t,G_1) + \vt_x(t,G_2)=0\, ,  
\label{g5} 
\end{eqnarray} 
the dot indicating the time derivative. Therefore the particle and energy inflows are equal to the outflows as expected. 

Let us consider now the operation of time-reversal. It is well known that this operation is implemented by an anti-unitary 
operator $T$, acting in the state space $\H$ of the system. We recall \cite{TR} that 
\begin{equation}
 T\, j_x(t,x)\, T^{-1} = -j_x(-t,x)   
\label{g11}
\end{equation} 
and assume that the dynamics preserves time-reversal invariance, namely 
\begin{equation}
T\, H\, T^{-1} = H \, . 
\label{ng1}
\end{equation}
Now, let $\Omega \in \H$ be any time-translation invariant state of the system with non-vanishing particle flow 
between the two reservoirs 
\begin{equation}
\langle j_x(t,x) \rangle_{{}_\Omega} \equiv (\Omega\, ,\, j_x(t,x) \Omega) \not = 0 \, ,  
\label{g10}
\end{equation}
where $(\cdot\, ,\, \cdot )$ is the scalar product in $\H$.
Taking the expectation value of (\ref{g11}) one has 
\begin{equation}
\langle T\, j_x(t,x)\, T^{-1}\rangle_{{}_\Omega} = -\langle j_x(-t,x)\rangle_{{}_\Omega}\, ,  
\label{g12}
\end{equation}
which, combined with the fact that $\langle j_x(t,x)\rangle_{{}_\Omega}$ is actually $t$-independent 
due to the time-translation invariance of $\Omega$, implies that 
\begin{equation}
T\, \Omega \not = \Omega \, . 
\label{g13}
\end{equation} 
We conclude that time-reversal is broken in the state $\Omega \in \H$ 
in spite of the fact that the dynamics is time-reversal invariant (\ref{ng1}). This result can be compactly 
formulated as follows. 
\medskip 

{\bf Proposition:} Any state, which is invariant under time translations and 
generates a non-vanishing expectation value for the particle current, 
breaks time-reversal symmetry. 
\medskip 

\noindent This is a genuine quantum field theory phenomenon of spontaneous breaking of a discrete 
symmetry, whose order parameter is $\langle j_x(t,x)\rangle_{{}_\Omega}$. 

The state $\Omega$ has another remarkable feature. Let 
$q_x(t, G_i)$ be the heat current flowing through the gate $G_i$. Since the value of the chemical potential 
in $G_i$ is $\mu_i$, one has \cite{Callen}  
\begin{equation}
q_x(t,G_i) = \vt_x (t,G_i) - \mu_i j_x(t,G_i)\, .  
\label{g6} 
\end{equation} 
Notice that we adopt here only the value of the heat current in the gates $G_i$. The point is that the 
heat current in the interaction domain $\D$ is not known, because the temperature and the chemical potential 
are not defined in this region. In order to introduce the concept of local parameters $\beta(x)$ and $\mu(x)$ for 
$x\in \D$ one needs further model dependent assumptions, which are not relevant and not needed for our construction.  

From (\ref{g6}) it follows that for $\mu_1\not = \mu_2$ the heat flow through $G_1$ differs from that through 
$G_2$. In fact, 
\begin{equation}
\dQ \equiv -q_x(t,G_2)-q_x(t,G_1) = (\mu_2-\mu_1) j_x(t,G_1) \not=0\, . 
\label{g7}
\end{equation}
Now we recall that the total energy $H$ of the system has two different components - {\it heat} and {\it chemical} energies. 
Since $H$ is conserved, (\ref{g7}) implies that these components are not separately conserved and can be converted 
one into the other. This process, discovered in \cite{MSS-14}, depends on the state $\Omega$ and more precisely on 
the expectation value 
\begin{equation}
\langle \dQ \rangle_{{}_\Omega} = (\Omega\, ,\, \dQ \Omega) \, . 
\label{g8}
\end{equation}
Chemical energy is converted to heat energy 
if $\langle \dQ \rangle_{{}_\Omega} >0$. The opposite process takes place for $\langle \dQ \rangle_{{}_\Omega} <0$ and energy 
transmutation is absent only if $\langle \dQ \rangle_{{}_\Omega} =0$. It is worth stressing that there is {\it no dissipation}  
during the energy conversion. 

Summarising, the physical consequences of particle and energy conservation in any state $\Omega$, 
satisfying (\ref{g10}), are: 
\medskip 

(i) spontaneous breaking of time-reversal; 

(ii) conversion of heat to chemical energy or vice versa. 
\medskip 

\noindent It is worth stressing that these deeply related features follow exclusively from 
symmetry considerations and do {\it not} depend on the interaction taking place in $\D$. 
In this sense they are {\it universal} and hold for all systems of the type shown in Fig. \ref{fig1}. 
\bigskip 

\section{Entropy production}

The properties (i)-(ii), established in the previous section, have relevant physical implications. Among others, they provide a simple and remarkable 
mechanism for non-vanishing entropy production $\dS$ in absence of dissipation. Our main goal in what follows will be to study 
this aspect in detail. According to the second law of classical thermodynamics $\dS \geq 0$. 
In the quantum case $\dS$ is an operator, which has the following form \cite{Callen} 
\begin{equation}
\dS = 
-\beta_1\, q_x(t,G_1)  - \beta_2\, q_x(t,G_2) 
\label{g14}
\end{equation}
in terms of the heat currents flowing through the gates $G_i$. In order to investigate the properties of this operator, one needs  
a microscopic approach \cite{MSS-17, MSS-18} to the particle transport in the system. It is based on the observation that (due to particle number conservation), there are three fundamental elementary processes which take place in the system in Fig. \ref{fig1}:  
\medskip 

(a) emission and reabsorption of any number of particles from the same reservoir;  

(b) emission of $n$ particles from the ``hot" reservoir $R_2$ and their absorption by the ``cold" one $R_1$;

(c) emission of $n$ particles from the ``cold" reservoir $R_1$ and their absorption by the ``hot" one $R_2$. 
\medskip 

\noindent Let us denote by $p_0$, $p_n$ and $p_{-n}$ with $n=1,2,...$ the probabilities of the events (a), (b) and (c) respectively. 
A substantial difference with respect to the classical case is that in the quantum world the processes of the type (c) have a priori non-vanishing 
probabilities. At this stage it is useful to introduce the sequences 
\begin{equation}
P=\{p_n\, :\, n=0,\pm1, \pm 2,...\}\, , \qquad \Sigma =\{\sigma_n\, :\, n=0,\pm1, \pm 2,...\}\, ,
\label{g15}
\end{equation}
where $\sigma_n$ is the entropy production associated with the process with probability $p_n$. 
It is natural to expect that $\sigma_0=0$, $\{\sigma_n>0,\, n=1,2,...\}$ and 
$\{\sigma_n<0,\, n=-1,-2,...\}$. 

Summarising, both processes with positive and negative entropy production occur at the microscopic level. It is clear that 
the sequences (\ref{g15}) fully codify the entropy production in the system.  So, the problem is to determine $P$ and $\Sigma$. 
We will show below that both $P$ and $\Sigma$ can 
be extracted from the probability distribution $\varrho [\dS]$ generated by the correlation functions 
\begin{equation} 
\M_n[\dS] = \langle  \dS (t_1) \cdots \dS(t_n)  \rangle_{{}_{\Omega}} \, . 
\label{g16}
\end{equation}
The based observation now is that $\M_n[\dS]$ are the moments of $\varrho [\dS]$, namely 
\begin{equation} 
\M_n[\dS]  = \int_{\cal D} \rd \sigma\, \sigma^n \varrho[\dS] (\sigma) \, , 
\label{g17}
\end{equation} 
where $\cal D$ is the range of entropy production. The strategy at this point consists of three steps: 
\begin{itemize}

\item derivation of the correlation functions $\M_n[\dS]$ given by (\ref{g16}); 

\item reconstruction of the distribution $\varrho [\dS]$ from its moments $\M_n[\dS]$ (also known as moment problem \cite{ST-70}); 

\item derivation of the sequences $P$ and $\Sigma$ from $\varrho [\dS]$. 

\end{itemize} 

\noindent Although conceptually very clear, the first two steps are practically quite involved. They include the derivation 
of the infinite sequence of correlation functions (\ref{g16}), followed by the solution of the moment problem. Nevertheless we 
give in the next section two examples where this program can be taken to the end. We describe one fermionic and one bosonic 
model, where the interaction domain $\D$ is reduced to a point $x=0$. We show that with such point-like interaction one can 
determine $\varrho [\dS]$ and consequently $P$ and $\Sigma$ in exact and explicit form and discuss the physical 
implications of the solution. 

\section{Fermionic/bosonic Schr\"odinger junction} 

\begin{figure}[h]
\begin{center}
\begin{picture}(600,22)(43,325) 
\includegraphics[scale=0.87]{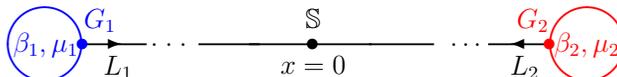}
\end{picture} 
\end{center}
\caption{System with point-like interaction $\S$ at $x=0$.} 
\label{fig2}
\end{figure}

We analyse in this section two examples with point-like interaction described by a unitary scattering matrix $\S$ 
as shown in Fig. \ref{fig2}. The dynamics along the leads $L_i$ with coordinates $\{(x,i) \, :\, x<0\, , i=1,2\}$, where $|x|$ is the 
distance from the interaction and $i$ labels the lead, is defined by the Schr\"odinger equation 
\begin{equation}
\left (\ri \prt_t +\frac{1}{2m} \prt_x^2\right )\psi (t,x,i) = 0\, , 
\label{e1}
\end{equation}
where $\psi$ is a quantum field satisfying the equal-time canonical (anti)commutation relations 
\begin{equation}
[\psi (t,x,i)\, ,\, \psi^*(t,y,j)]_{{}_\pm} = \delta_{ij}\, \delta(x-y) 
\label{e2}
\end{equation}
and $^*$ stands for Hermitian conjugation. The interaction is fixed by the boundary condition 
\begin{equation}
\lim_{x\to 0^-}\sum_{j=1}^2 \left [\lambda (\II-\UU)_{ij} +\ri (\II+\UU)_{ij}\prt_x \right ] \psi (t,x,j) = 0\, ,
\label{e3}
\end{equation}
where $\lambda$ is a free parameter with dimension of mass and $\UU$ is an arbitrary unitary matrix 
$\UU\in U(2)$. This is \cite{ks-00} the {\it most general} boundary condition ensuring the {\it self-adjointness} of the Hamiltonian. 
In momentum space the associated scattering matrix is given by \cite{ks-00, h-00} 
\begin{equation}
\S(k) = -\frac{[\lambda (\II - \UU) - k(\II+\UU )]}{[\lambda (\II - \UU) + k(\II+\UU)]} \, , 
\label{e4}
\end{equation}
where the matrices in the numerator and denominator commute. One can easily verify that this scattering matrix is 
unitary $\S(k) \S^*(k) = \II$ and satisfies Hermitian analyticity $\S^*(k) = \S(-k)$. The dynamics fixed by (\ref{e1}-\ref{e3}) is invariant under 
time-reversal if and only if $\UU$ and consequently $\S(k)$ are {\it symmetric}. 

We will show below that the above system, called in what follows Schr\"odinger junction, nicely illustrates the 
program from the previous section and works simultaneously for both Fermi $(+)$ and Bose $(-)$ statistics. For simplicity we assume 
in what follows that $\S(k)$ has no bound states, referring for the general case to \cite{MSS-17an}. Then, the general solution of (\ref{e1}-\ref{e3}) 
is given by 
\begin{equation} 
\psi (t,x,i)  = \sum_{j=1}^2 \int_{0}^{\infty} \frac{dk}{2\pi } 
\e^{-\ri \omega (k)t}\, \left [ \e^{-\ri k x}\, \delta_{ij} +\e^{\ri k x}\, \S_{ij}^*(k)\right ] a_j (k) \, , 
\label{e5} 
\end{equation} 
where $\omega(k) = \frac {k^2}{2m}$ is the dispersion relation 
and the oscillators 
\begin{equation}
\{a_i(k),\, a^*_i(k),\, :\, ,k>0,\, i=1,2\} 
\label{e51}
\end{equation}
generate a standard canonical (anti)commutation relation algebra $\A_\pm$ 
\begin{equation}
[a_i(k)\, ,\, a_j^*(p)]_{{}_\pm} = \delta_{ij} \, 2\pi \delta(k-p)\, . 
\label{e7}
\end{equation} 
Notice that (\ref{e51}) annihilate and create only in-coming particles because $k>0$. The contribution 
of the out-going excitations is generated by the second term in the integrand of (\ref{e5}), which involves the 
scattering matrix $\S(k)$. 

The basic physical observables, we are interested in, are 
\begin{eqnarray}
j_x(t,x,i)= \frac{\ri }{2m} \left [ \psi^* (\partial_x\psi ) - 
(\partial_x\psi^*)\psi \right ]  (t,x,i)\, ,   
\label{curr} \\
\vartheta_x (t,x,i) = \frac{1}{4m} [\left (\partial_t \psi^* \right )\left (\partial_x \psi \right ) 
+ \left (\partial_x \psi^* \right )\left (\partial_t \psi \right ) \; 
\nonumber \\
-\left (\partial_t \partial_x \psi^* \right ) \psi - 
\psi^*\left (\partial_t \partial_x \psi \right ) ](t,x,i)\, ,  
\label{en1}
\end{eqnarray}
\begin{equation}
\dS (t,x) = - \sum_{i=1}^2 \beta_i\, q (t,x,i) \, ,\qquad  q_x(t,x,i) = \vartheta_x (t,x,i) - \mu_i j_x(t,x,i)\, . 
\label{dS}
\end{equation} 
In order to compute the correlation functions of the entropy production operator $\dS(t,x)$ one must fix 
a representation of the oscillator algebras $\A_\pm$. We choose the Landauer-B\"uttiker (LB) 
representation \cite{L-57, B-86} $\{\H_{{}_{\rm LB}},\, \Omega_{{}_{\rm LB}}\}$, 
which represents a non-equilibrium generalisation of the 
Gibbs representation \cite{BR} to the case of systems which exchange particles 
and energy with more then one heat reservoir. Adopting the field-theoretical construction \cite{M-11}
of $\{\H_{{}_{\rm LB}},\, \Omega_{{}_{\rm LB}}\}$, one can derive in explicit form the expectation values 
\begin{equation}
\langle a^*_{l_1}(k_1) a_{m_1}(p_1)\cdots a^*_{l_n}(k_n) a_{m_n}(p_n)\rlb^\pm \, , \qquad k_i, p_i >0\, ,   
\label{e8}
\end{equation}
in the LB {\it steady} state $\Omega_{{}_{\rm LB}}$. It is convenient for this purpose to introduce the matrix 
\begin{equation} 
\mM^\pm_{ij} = 
\begin{cases} 
2\pi \delta (k_i-p_j)\delta_{l_im_j} d^\pm_{l_i}[\omega(k_i)]\, ,\qquad \qquad \qquad i\leq j\, , \\
\mp 2\pi \delta (k_i-p_j)\delta_{l_im_j}\left (1\mp d^\pm_{l_i}[\omega(k_i)]\right )\, ,\qquad i > j\, ,\\ 
\end{cases} 
\label{e9}
\end{equation} 
where 
\begin{equation}
d^\pm_l(\omega ) = \frac{1}{\e^{\beta_l (\omega - \mu_l)} \pm 1}\, ,   \qquad \quad ({\rm for\; bosons}\; \mu_l<0)
\label{e10}
\end{equation}  
are the Fermi/Bose distribution of the reservoir $R_l$. Then one has \cite{M-11}
\begin{equation}
\langle a^*_{l_1}(k_1) a_{m_1}(p_1)\cdots a^*_{l_n}(k_n) a_{m_n}(p_n)\rlb^\pm  = 
\begin{cases} 
{\rm \bf det}\, [\mM^+]\, ,  \quad k_i, p_i >0\, , \\ 
{\rm \bf perm}\, [\mM^-]\, ,\quad k_i, p_i >0\, ,\\ 
\end{cases} 
\label{e11}
\end{equation} 
where ${\rm \bf det}$ and ${\rm \bf perm}$ indicate the determinant and the permanent of the corresponding matrices. 
It is perhaps useful to recall that 
\begin{equation}
{\rm \bf perm}\, [\mM]= \sum_{\sigma_i \in \cP_n} \prod_{i=1}^n \mM_{i \sigma_i} \, , \qquad 
\cP_n - {\rm set\; of\; all\; permutations\; of\; } n\; {\rm elements} \, . 
\label{e12} 
\end{equation}

Equations (\ref{e9}-\ref{e11}) are the basic ingredients for deriving the correlation functions in the LB representation. 
Our first step in this direction will be to compute some one-point functions. In particular, we will 
check that time-reversal is spontaneously broken, namely 
that $T\Omega_{{}_{\rm LB}} \not= \Omega_{{}_{\rm LB}}$. Following the argument in section 1, it is enough to control the 
mean value of the particle current (\ref{curr}). One finds 
\begin{equation}
\langle j(t,x,i)\rlb^\pm = \int_{0}^\infty \frac{\rd \omega}{2\pi} \sum_{l=1}^2 \left [\delta_{il} - |\S_{il}(\sqrt {2m\omega})|^2\right ] 
d_l^\pm (\omega) \not = 0\, , 
\label{sb}
\end{equation}
which confirms the spontaneous $T$-breaking. 

{}For the mean value of the entropy production one has 
\begin{equation}
\langle \dS(t,x)\rlb^\pm   = \int_{0}^\infty \frac{\rd \omega}{2\pi} |\S_{12}(\sqrt {2m\omega})|^2 [\gamma_2(\omega)-\gamma_1(\omega)] 
[d_1^\pm (\omega)-d_2^\pm(\omega)] \, \geq 0\, , 
\label{e13}
\end{equation} 
where 
\begin{equation}
\gamma_i(\omega) \equiv \beta_i(\omega -\mu_i)\, .  
\end{equation}
The positivity follows directly from the fact that the two square brackets 
in the integrand have always the same sign. The bound (\ref{e13}) is a special case of the general result of Nenciu \cite{N-07} 
for systems in the LB state and in contact with arbitrary number of heat reservoirs. 

We turn now to the $n$-point correlation functions 
\begin{equation}
\langle \dS(t_1,x_1) \cdots \dS(t_n,x_n)\rlb^\pm \, ,  
\label{e14}
\end{equation}
which, due to the total energy conservation, depend only on the time differences $\{\wt_i \equiv t_i - t_{i+1}\, :\, i=1,...,n-1\}$. 
This fact allows to introduce for $n\geq 2$ the frequency $\nu$ via the Fourier transform 
\begin{eqnarray}
w_n^\pm[\dS](x_1,...,x_n;\nu) = \qquad \quad \qquad \qquad \qquad 
\nonumber \\
\int_{-\infty}^{\infty} \rd \wt_1 \cdots   \int_{-\infty}^{\infty} \rd \wt_{n-1} 
\e^{-\ri \nu (\wt_1+\cdots \wt_{n-1})} \langle \dS(t_1,x_1) \cdots \dS(t_n,x_n)\rlb^\pm \, .
\label{e15}
\end{eqnarray} 
Following the classical studies \cite{ML-92}-\cite{MSS-15} in quantum noise and full 
counting statistics, we perform the zero-frequency limit 
in which the quantum fluctuations are integrated over the whole time axes. It turns out 
\cite{MSS-17, MSS-18} that in this limit 
the position dependence in (\ref{e15}) drops out and one finds 
\begin{equation}
\lim_{ \nu \to 0^+} w_n^\pm[\dS](x_1,...,x_n;\nu)  = \int_0^\infty \frac{\rd \omega}{2\pi}\, \M^\pm_n[\dS] (\omega)\, ,
\label{e16}
\end{equation}
where $\M^\pm_n[\dS]$ are precisely the moments (\ref{g17}) of the probability distribution 
$\varrho^\pm_n[\dS]$ we are looking for. Using (\ref{e9}-\ref{e11}), one gets 
\begin{equation}
\M^\pm_n[\dS] = 
\begin{cases}
\gamma_{21}^n(\omega )\, {\rm \bf det}[\D^+(\omega;l_1,...,l_n)]\, ,\\
\gamma_{21}^n(\omega )\, {\rm \bf perm}[\D^-(\omega;l_1,...,l_n)]\, ,\\ 
\end{cases} 
\label{e17}
\end{equation} 
where 
\begin{equation}
\gamma_{ij}(\omega ) \equiv \gamma_i(\omega )-\gamma_j(\omega) = (\beta_i-\beta_j)\omega -(\beta_i\mu_i-\beta_j\mu_j)\, , 
\label{e18}
\end{equation}
is a basic dimensionless parameter defining the entropy generated by transporting a particle 
with energy $\omega$ from the reservoir $R_i$ to $R_j$. Moreover the $\D^\pm$-matrices are given by 
\begin{equation} 
\D^\pm_{ij} (\omega;l_1,...,l_n)= 
\begin{cases} 
\mJ_{l_jl_i}(\omega) d^\pm_{l_j}(\omega)\, ,\qquad \qquad \qquad i\leq j\, , \\
\mp \mJ_{l_jl_i}(\omega )\left [1\mp d^\pm_{l_j}(\omega)\right ]\, ,\qquad i > j\, ,\\ 
\end{cases} 
\label{e19}
\end{equation} 
with 
\begin{eqnarray}
\mJ_{11}(\omega )=-\mJ_{22}(\omega ) = |\S_{12}(\sqrt {2 m \omega})|^2 \equiv \tau (\omega )\, , \; \; \; 
\label{e20}\\
\mJ_{12}(\omega )={\overline \mJ}_{21}(\omega )=-\S_{11}(\sqrt {2 m \omega})\, {\overline\S}_{12}(\sqrt {2 m \omega})\, , 
\label{e21}
\end{eqnarray} 
where $\tau (\omega)$ is the {\it transmission probability}.  For $\tau(\omega) = 0$ the leads are isolated and the system is in equilibrium.  

At this stage, using (\ref{e17}-\ref{e21}) one can derive the inequality 
\begin{equation}
\M^\pm_n[\dS] \geq 0 \, , \qquad n=,1,2,...
\label{e22}
\end{equation}
which is one of our main results. The rigorous proof of the bound (\ref{e22}) can be found in \cite{MSS-17} for 
fermions and in \cite{MSS-18} for bosons. Referring to these papers for the details, we will give below an 
alternative intuitive explanation, which has the advantage of providing a simple physical interpretation in terms 
of the sequence of probabilities $P$ in (\ref{g15}). In order to derive $P$ we introduce the moment 
generating function \cite{ST-70} 
\begin{equation}
\chi^\pm [\dS](\lambda ) = \sum_{n=0}^\infty \frac{(\ri \lambda)^n}{n!} \M^\pm_n[\dS] \, , 
\label{e23}
\end{equation} 
where the $\omega$-dependence is implicit. After some algebra one finds 
\begin{eqnarray}
\chi^+[\dS](\lambda ) = 1+ \ri c_1^+ \sqrt {\tau} \sin(\lambda \gamma_{21}\sqrt {\tau})+
c_2^+\left [ \cos(\lambda \gamma_{21}\sqrt {\tau}) - 1\right ]\, ,
\label{e24} \\
\chi^-[\dS](\lambda ) = \frac{1}{1- \ri c_1^- \sqrt {\tau} \sin(\lambda \gamma_{21}\sqrt {\tau})-
c_2^-\left [ \cos(\lambda  \gamma_{21} \sqrt {\tau}) - 1\right ]}\, ,\; \, 
\label{e25}
\end{eqnarray}
where $c_i^\pm$ are expresses via the reservoir distributions $d_i^\pm$ as follows 
\begin{equation}
c_1^\pm \equiv d^\pm_1 - d^\pm_2  \, , \qquad 
c_2^\pm \equiv d_1^\pm + d_2^\pm \mp 2 d_1^\pm d_2^\pm \, . 
\label{e26}
\end{equation} 

The final step towards the entropy production distributions $\varrho^\pm[\dS]$ is the Fourier transform 
\begin{equation}
\varrho^\pm [\dS](\sigma )= \int_{0}^\infty \frac{\rd \lambda}{2\pi} \e^{-\ri \lambda \sigma }\chi^\pm [\dS](\lambda )\, . 
\label{e27}
\end{equation}
Since according to (\ref{e24},\ref{e25}) $\chi^\pm [\dS]$ are periodic functions in $\lambda $ with period $2\pi/\sqrt{\tau}$, 
the Fourier transform is a superposition of $\delta$-functions - the so called ``Dirac comb". 
The fermionic comb has only three ``teeth"  
\begin{equation}
\varrho^+[\dS](\sigma ) = \sum_{k=-1}^1 p^+_k\, \delta(\sigma - k\gamma_{21}\sqrt \tau) \, , 
\label{e28}
\end{equation}
because only single particle processes are allowed by Pauli's principle since the energy $\omega$ 
is fixed and there is no degeneracy in spin and momentum in the Schr\"odinger junction. 
In (\ref{e28}) 
\begin{equation}
p^+_{\pm 1} = \frac{1}{2}\left (c^+_2 \mp c^+_1\sqrt {\tau}\right )\, ,\qquad  p^+_0 = 1-c^+_2\, , 
\label{e29}
\end{equation} 
and 
\begin{equation}
\sigma_{\pm 1} = \pm \gamma_{21}\sqrt \tau \, , \qquad \sigma_0 = 0\, , 
\label{e30}
\end{equation} 
are precisely the elements of the sequences $P^+$ and $\Sigma^+$ we are looking for. In fact one can easily verify that 
\begin{equation}
p^+_k \geq 0\, , \qquad \sum_{k=-1}^1 p^+_k = 1\, . 
\label{e31}
\end{equation}

The bosonic case is technically more involved since the comb has infinite ``teeth" 
\begin{equation}
\varrho^-[\dS](\sigma ) = \sum_{k=-\infty}^\infty p^-_k\, \delta(\sigma - k\gamma_{21}\sqrt \tau) \, , 
\label{e32}
\end{equation}
because multi-particle processes are allowed. The computation, performed in \cite{MSS-18}, gives   
\begin{equation}
p^-_{\pm n} = \frac{(c^-_2\pm c^-_1\sqrt \tau)^n}{2^n(1+c^-_2)^{n+1}}\, {}_2F_1 \left [\frac{1+n}{2}, \frac{2+n}{2}, n+1,
\frac{(c^-_2)^2 - \tau (c^-_1)^2}{(1+c^-_2)^2}\right ] \, , 
\label{e33}
\end{equation} 
\begin{equation}
\sigma_{\pm n} = \pm n \gamma_{21}\sqrt \tau \, , \qquad n=0,\pm 1,\pm2,... 
\label{e331}
\end{equation}
where ${}_2F_1$ is the Gauss hypergeometric function. Equations (\ref{e33},\ref{e331}) define the sequences 
$P^-$ and $\Sigma^-$ in the bosonic case. In analogy with (\ref{e31}) one has \cite{MSS-18}  
\begin{equation}
p^-_k \geq 0\, , \qquad  \sum_{k=-\infty}^\infty p^-_k = 1\, .  
\label{e34}
\end{equation} 

\begin{figure}[h]
\begin{center}
\begin{picture}(15,90)(170,30) 
\includegraphics[scale=0.315]{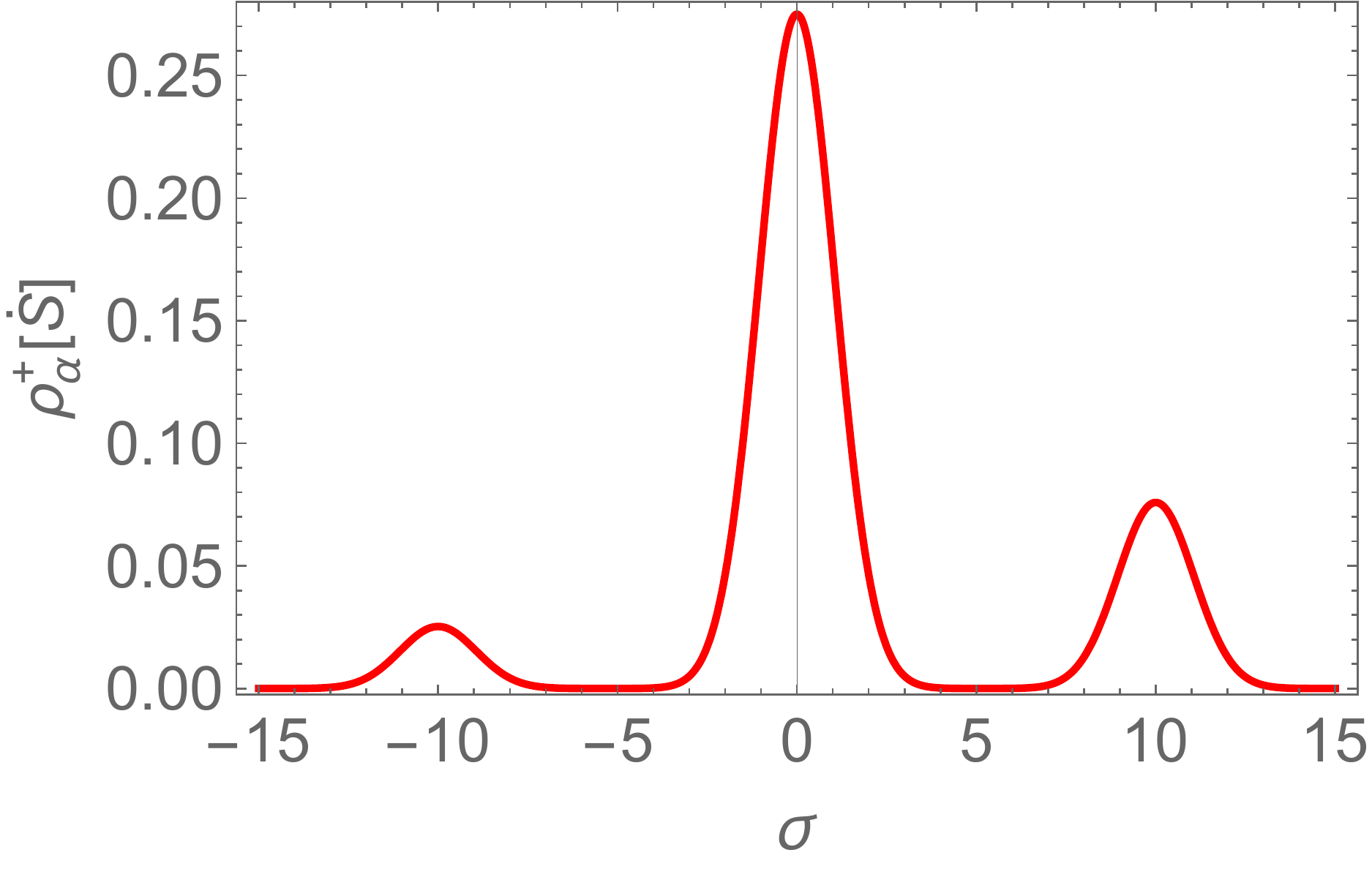} 
\hskip 0.2 truecm
\includegraphics[scale=0.305]{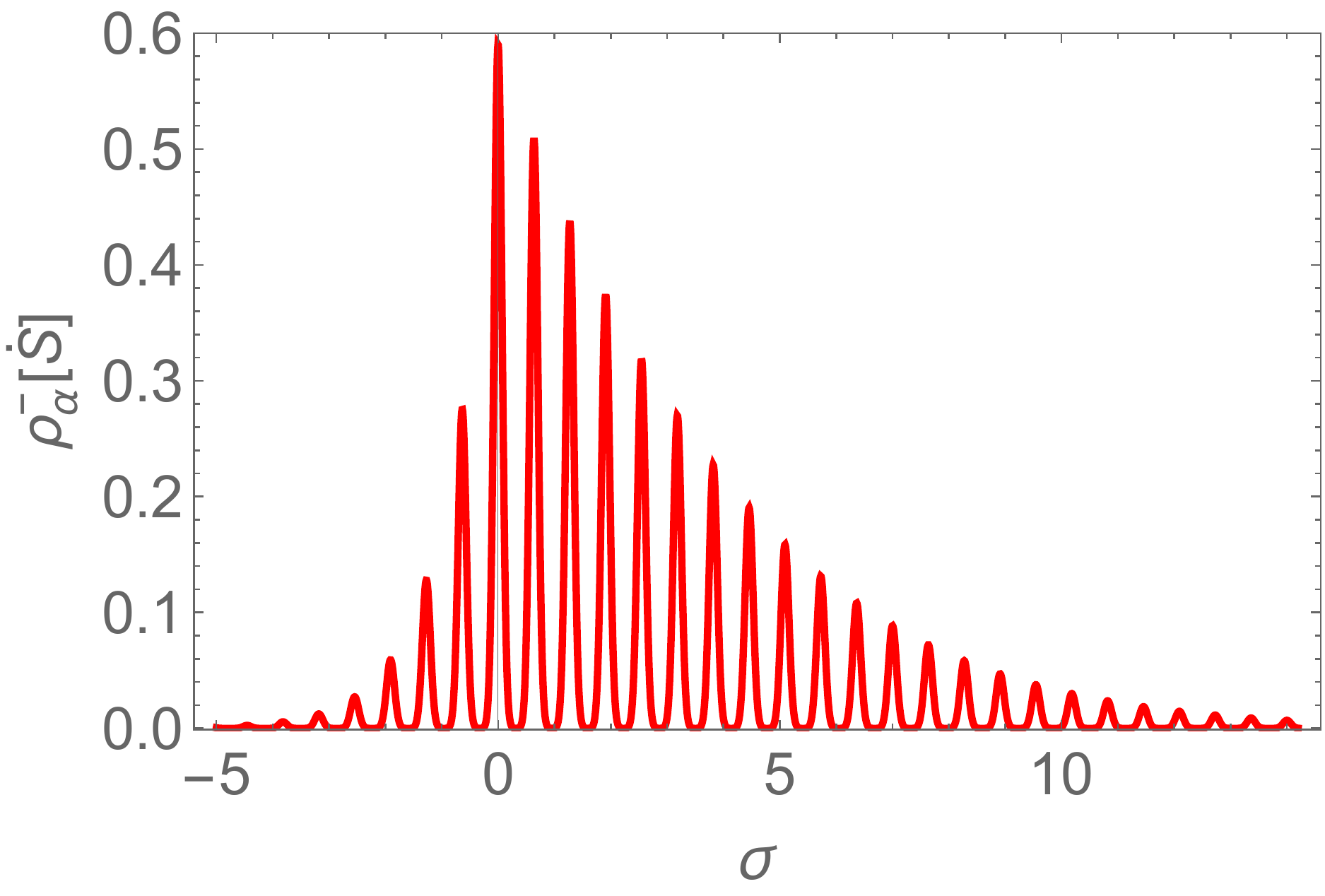}
\end{picture} 
\end{center}
\caption{Fermionic and bosonic entropy production distributions.} 
\label{fig3}
\end{figure}

Summarising, (\ref{e31},\ref{e34}) imply that both $\varrho^\pm [\dS]$ are well defined probability distributions. 
Equations (\ref{e29},\ref{e33}) provide in explicit form the probabilities $p_k^\pm$ for the basic emission and absorption processes in 
terms of the Dirac/Bose distributions $d_i^\pm$ and the interaction $\tau$. All quantum fluctuations in the zero frequency 
limit are taken into account. The relative simplicity of $p_k^+$ with respect to $p_k^-$ is a consequence of the Pauli principle. 
In order to compare $p_k^\pm$ for different $k$ and to get a more precise idea about the distributions $\varrho^\pm[\dS]$ it 
is convenient to introduce the smeared versions $\varrho_\alpha^\pm[\dS]$ by the substitution 
\begin{equation}
\delta (\sigma ) \longmapsto \delta_\alpha (\sigma ) \equiv \frac{\alpha}{\sqrt \pi}\, \e^{-\alpha^2 \sigma^2}\, , \qquad \alpha>0\, , 
\label{e35}
\end{equation}
in (\ref{e28},\ref{e32}). By construction $\varrho_\alpha^\pm[\dS] \rightarrow \varrho^\pm[\dS]$ for $\alpha \to \infty$ in the 
sense of generalised functions. The advantage of $\varrho_\alpha^\pm[\dS]$ is that they are not singular and can be easily plotted 
for some value of $\alpha$ and the parameters $\beta_i$, $\mu_i$ and $\tau$. Typical plots are shown in Fig. \ref{fig3}, where the left 
and right panel display the fermionic and bosonic distribution respectively. 

The relative height of the peaks in Fig. \ref{fig3} allows to compare the different 
probabilities of the fundamental emission-absorption processes described in points (a)-(c) in section 3. 
We see that the predominant process is the emission and reabsorption by the same reservoir 
with vanishing entropy production $\sigma =0$. For $\sigma\not=0$ the peaks are symmetric 
with respect to the origin and the right ones ($\sigma>0$) 
always dominate the left ones ($\sigma<0$). This observation provides the physical explanation for the positivity bound (\ref{e22}) 
on the moments $\M_n^\pm [\dS]$. 

The knowledge of the sequences $P^\pm$ and $\Sigma^\pm$ in explicit form 
(\ref{e29},\ref{e30},\ref{e33},\ref{e331}) has 
relevant physical applications. Using this microscopic information one can establish the 
fluctuation relations \cite{C-99}-\cite{ES-02}  
governing the entropy production in the LB state. Moreover, one can introduce a concept of efficiency of the 
quantum transport, which goes beyond the meal value description and takes into account all 
quantum fluctuations. For more details about these applications we refer the reader to \cite{MSS-17, MSS-18}. 

In conclusion, for non-equilibrium quantum systems in the LB state microscopic processes with 
negative entropy production occur with some non-vanishing probability. For each such process however, there exists 
a more probable one with the opposite positive entropy production, which dominates. This feature can be 
interpreted in our context as a {\it quantum} version of the second law of thermodynamics.

\section{Conclusions}

The above study develops a field theoretic approach to  
the entropy production in non-equilibrium quantum systems 
in a state $\Omega$, which breaks spontaneously time-reversal invariance. 
The parameter which controls this phenomenon is the expectation value of the 
particle current in $\Omega$. The mechanism allows for non-trivial entropy 
production even in absence of dissipation. The basic idea of the framework is 
to derive and investigate the probability distribution 
$\varrho[\dS]$, generated by the $n$-point correlation functions of the entropy 
production operator $\dS$ in the state $\Omega$. One can extract from $\varrho[\dS]$ the 
sequence $P$ of probabilities, associated with the fundamental  
processes of emission and absorption of particles from the heat reservoirs, 
driving the system away from equilibrium. In this way one obtains a microscopic 
picture of the quantum transport and entropy production, which takes into account 
all quantum fluctuations. These general ideas have been illustrated in the paper 
on the example of two exactly solvable models - the fermionic and bosonic 
Schr\"odinger junctions with point-like interaction. We show that in these two 
cases one can derive in exact and closed form the probability distribution 
$\varrho[\dS]$ and prove that all its moments $\M_n[\dS]$ are non-negative in the zero 
frequency limit, which provides a bridge with the second law of non-equilibrium 
classical thermodynamics. 

Our investigation demonstrates that the entropy production operator plays 
a fundamental role in non-equilibrium quantum physics. For this reason it will 
be important to test the above ideas in other models and within alternative frameworks. 
Among others, non-equilibrium conformal field theory \cite{BD-15}-\cite{MS-13}, 
generalised quantum hydrodynamics \cite{F-17}-\cite{D-19} 
and the theory of periodically driven quantum systems \cite{B-20} could be adopted. 
Studies in this direction will certainly contribute 
for better understanding the fascinating properties of the quantum world away from equilibrium. 
\bigskip 

{\bf Acknowledgements:} One of us (P.S.) would like to thank  the organisers Prof. B. Dragovich 
and Dr. M. Vojinovic for their kind  invitation to present our results at MPHYS10, Belgrade, Sept. 2019. 
The scientific quality of the meeting as well as the perfect organisation have been particularly appreciated.

\end{document}